 \documentstyle[prb,aps,epsf,floats,twoside]{revtex}

\begin{document}

  \twocolumn[\hsize\textwidth\columnwidth\hsize\csname
  @twocolumnfalse\endcsname

\draft

\title{Weak Antilocalization and Spin Precession in Quantum Wells}

\author{W.Knap,        \\        C.        Skierbiszewski\cite{Auth1},
A.Zduniak\cite{Auth2}, E. Litwin-Staszewska\cite{Auth1}, D.Bertho,  F.
Kobbi, and J. L. Robert}

\address{Groupe d'Etude des Semiconducteurs, Universite Montpellier II
C.N.R.S. URA 357,  Place E. Bataillon,  34095 - Montpellier-Cedex  05,
FRANCE}

\author{G. E. Pikus}

\address{A.F. Ioffe Physicotechnical Institute 194021 St  Petersbourg,
RUSSIA}

\author{F. G. Pikus}

\address{Department  of  Physics,  University  of  California at Santa
Barbara, Santa Barbara, CA 93106, USA}

\author{S. V. Iordanskii}

\address{Landau  Institute  for  Theoretical  Physics,  117940 Moscow,
RUSSIA}

\author{V. Mosser}

\address{Schlumberger E.  T. L.  50, Avenue  Jean Jaures,  BP. 620-05,
92542 - Montrouge, FRANCE}

\author{K. Zekentes}

\address{Forth Institute of Electronic  Structure and Laser, P.O.  Box
1527, Heraklion, 71110 - Crete, GREECE}

\author{and Yu. B. Lyanda-Geller}

\address{Beckman    Institute,     University    of     Illinois    at
Urbana-Champaign, Urbana, IL 61801}

\date{\today}

\maketitle

\begin{abstract}

The  results  of  magnetoconductivity  measurements  in GaInAs quantum
wells are presented. The  observed magnetoconductivity appears due  to
the quantum interference, which lead to the weak localization  effect.
It  is  established  that  the  details  of  the weak localization are
controlled by  the spin  splitting of  electron spectra.  A theory  is
developed which takes into account  both linear and cubic in  electron
wave vector terms in  spin splitting, which arise  due to the lack  of
inversion center  in the  crystal, as  well as  the linear terms which
appear when  the well  itself is  asymmetric. It  is established that,
unlike spin  relaxation rate,  contributions of  different terms  into
magnetoconductivity are not additive.  It is demonstrated that  in the
interval     of     electron     densities     under     investigation
{\bf(}$(0.98-1.85)\cdot  10^{12}\   {\rm  cm}^{-2}${\bf)}   all  three
contribution  are  comparable  and  have  to  be taken into account to
achieve  a  good  agreement  between  the  theory  and experiment. The
results obtained from comparison of the experiment and the theory have
allowed us to determine  what mechanisms dominate the  spin relaxation
in quantum wells and to improve the accuracy of determination of  spin
splitting  parameters  in  ${\rm  A}_3  {\rm  B}_5$  crystals and $2D$
structures.

\end{abstract}


  \vskip 2pc ] 

\narrowtext

\section{Introduction}
\label{sec:Intro}

The effect of  the weak localization  in metals and  semiconductors is
caused by the interference of  two electron waves which are  scattered
by the same centers (defects or impurities) but propagate in  opposite
directions along the same closed trajectory, and, therefore, return to
the  origin  with  equal  phases.  This effect increases the effective
scattering  crossection,  and,  therefore,  leads  to  a suppression of
conductivity\cite{Abrahams79,Gorkov79,FootnoteA}. In a magnetic field,
the two waves propagating in  the opposite directions acquire a  phase
difference $2e\Phi/c$, where $\Phi$  is the magnetic flux  through the
area enclosed by the electron trajectory. This phase difference breaks
the constructive  interference and  restores the  conductivity to  the
value it would have without the quantum interference corrections.  This
is observed as an {\em increase} in conductivity with magnetic  field,
the    effect     known    as     {\em positive    magnetoconductivity} (PMC)
or {\em negative magnetoresistance}
\cite{Altshuler80,Hikami80}.

When spin  effects are  taken into  account, the  interference depends
significantly on the total spin of the two electron waves. The singlet
state with the total spin $J  = 0$ gives negative contribution to  the
conductivity (antilocalization effect). The triplet state with $J = 1$
gives positive  contribution to  the conductivity.  In the  absence of
spin relaxation the contribution of the singlet state is canceled  by
one of triplet states. As  a result, the magnetic field  dependence of
the  conductivity  is  the  same  as  for spinless particles. However,
strong spin  relaxation can  suppress the  triplet state  contribution
without changing that  of the singlet  state, hence the  total quantum
correction  may  become  positive.  The  interplay  between   negative
magnetoconductivity    at   low    fields   and    positive
magnetoconductivity at high fields can lead to appearance of a minimum
on  the  conductivity   --  magnetic  field   curve  (antilocalization
minimum).

It  was   shown  in   the  early   papers  by   Hikami,  Larkin,   and
Nagaoka\cite{Hikami80}   and   by   Altshuler,   Aronov,  Larkin,  and
Khmelnitskii\cite{Altshuler81} that the  behavior of the  conductivity
in weak magnetic  fields depends essentially  on the mechanism  of the
spin  relaxation.  Three  mechanisms  were  considered: Elliott-Yafet,
otherwise   known   as   skew   scattering  mechanism,  scattering  on
paramagnetic impurities, and  Dyakonov--Perel mechanism, which  arises
from  the  spin  splitting  of  carrier spectra in non-centrosymmetric
media.

Dyakonov--Perel mechanism is  dominant in most  ${\rm A}_3 {\rm  B}_5$
cubic semiconductors\cite{Pikus84}, with  the exception of  those with
narrow band gap  $E_g$ and large  spin-orbit splitting of  the valence
band $\Delta$  (for example,  InSb). The  same can  be said  about the
low-dimensional structures fabricated  from these materials.  Presence
of a  antilocalization minimum  on the  $\sigma(B)$ curves  in quantum
$2D$  systems  is  a  definite  sign that the dominant spin-relaxation
mechanism is the Dyakonov--Perel one. It is known\cite{Hikami80}  that
for the Elliott-Yafet mechanism in $2D$ structures the contribution of
the singlet  state with  $J =  0$ is  exactly canceled  by one of the
triplet states, the one with $J = 1$ and $J_z = 0$.

Unlike bulk crystals, where the spin splitting is proportional to  the
cube of the wave vector $k$, in $2D$ structures the splitting has also
terms linear in $k$  (this is also true  for strained ${\rm A}_3  {\rm
B}_5$ crystals  and for  hexagonal ${\rm  A}_2 {\rm  B}_6$ compounds).
Furthermore, there are two linear in $k$ contributions of  essentially
different  nature.  The  first  one,  which  arises  from  the lack of
inversion in the original crystal  (like the cubic term), is  known as
Dresselhaus term\cite{Dresselhaus55}, while  the second, Rashba  term,
is  caused  by  the  asymmetry  of  the quantum well or heterojunction
itself\cite{Bychkov84}.

The  direct  measurements  of  the  spin  splitting  using  the  Raman
scattering in GaAs/AlGaAs  quantum wells\cite{Jusserand95} have  shown
that, for electron densities  $N_s \sim 10^{12}\ {\rm  cm}^{-2}$, both
linear contributions  are comparable.  All three  terms give  additive
contributions to the spin relaxation rate. When only the cubic in  $k$
term is present, its effect on the PMC is determined only by the  spin
relaxation  rate,   similarly  to   the  two   other  spin  relaxation
mechanisms,     and     is     described     by    the    theory    of
Refs.~\onlinecite{Hikami80,Altshuler81}. In the presence of the linear
in $k$  term in  the spin  Hamiltonian it  is necessary  to take  into
account  the  correlations  between  the  motion  of  electrons in the
coordinate and spin spaces \cite{FootnoteB}. In the theory of coherent
phenomena these correlations were  first taken into account  using the
language    of    the     spin-dependent    vector    potential     in
Refs.~\onlinecite{Mathur92,Aronov93,Geller94}, where this concept  was
applied  to  consideration  of  spin-orbit  conductance   oscillations
\cite{Mathur92,Aronov93}  and  of  the  spin  orbit  effects  in   the
universal   conductance   fluctuation   and   persistent   current  in
rings\cite{Geller94}.    In    the    theory    of    the    anomalous
magnetoconductivity the  correlation between  motion in  real and spin
spaces was first taken into account in Ref.~\onlinecite{Iordanskii94}.
It was  shown that  when linear  and cubic  in $k$  terms are present,
their contributions to spin  phase-breaking are not additive.  Furthermore,
as it was demonstrated in Ref.~\onlinecite{Pikus95}, the contributions
of  Rashba  and  Dresselhaus  terms  are  also  nonadditive,  and  the
magnetoconductivity is determined not  by their sum, but  rather their
difference.
Similar effect, when
spin-orbit phase-breaking may become negligible due to the correlation
of
motion in real and spin spaces occurs in quasi-$1D$ case and
leads to two types of pronounced oscillations in the universal
conductance fluctuations in rings\cite{Nottingham95}.

First experimental studies of the PMC in quantum $2D$ structures  were
done in Refs.~\onlinecite{Poole82,Beresovetz84,Kawaguchi87}. Recently,
the  experimental  observations  of  very  pronounced  effects of spin
scattering  on  weak  localization  conductivity  corrections  in GaAs
\cite{Dresselhaus92,Hansen93} and InAs \cite{Chen93}  heterostructures
have been reported. Different  spin relaxation mechanisms are  invoked
to      explain      the      experimental      data.      In      the
Ref.~\onlinecite{Dresselhaus92} the spin relaxation is interpreted  in
the framework of the Dyakonov-Perel  mechanism based on the bulk  GaAs
Hamiltonian.  According  to  the  authors, agreement with experimental
data is  achieved if  one neglects  in the  spin-orbit Hamiltonian the
term linear in the in-plane wave vector. This is in contradiction with
the theoretical  predictions\cite{Pikus95} which  show that,  at least
for low carrier concentrations, the linear term should be the dominant
one. In Ref.~\onlinecite{Chen93} the same Dyakonov-Perel mechanism  is
used,    but    it    is    based    on    the    Rashba    term.   In
Ref.~\onlinecite{Hansen93} it is  assumed that the  dominant mechanism
of spin relaxation is  the Elliott-Yafet scattering one.  Recently, we
have reported  the measurements  of the  magnetoconductivity of GaInAs
quantum        wells\cite{Knap94}.         We        have         used
Altshuler-Aronov-Larkin-Khmelnitskii        (AALKh)        calculation
\cite{Altshuler81} of quantum corrections to conductivity to interpret
the experimental data. However, the spin relaxation time, which enters
the AALKh  expressions, was  calculated taking  into account  not only
linear but also cubic Dresselhaus terms of the spin splitting. We have
demonstrated that, using this simplified theoretical approach, one can obtain right order  of
magnitude for the experimentally observed spin relaxation rates.

In this work  we present detailed  experimental study of  the negative
magnetoconductivity  in  selectively  doped  GaInAs quantum wells with
different $2D$ carrier densities.  We interpret them in  the framework
of recently  developed theory  of the  anomalous magnetoconductivity in
quantum       wells        which       corrects        the       AALKh
approach\cite{Iordanskii94,Pikus95}.  Comparison  of  experiment   and
theory allows to determine  importance of both Dresselhaus  and Rashba
terms for 2D systems.

\section{Theory}
\label{sec:Theory}

\subsection{Spin relaxation}
\label{subsec:Spin}

The theory  of the  positive magnetoconductivity  (PMC) for structures
with spin splitting linear in  wave vector was described very  briefly
in  Refs.~\onlinecite{Iordanskii94,Pikus95}.   Below  we   present  an
outline of  this theory  in more  details. The  spin splitting  of the
conduction band in cubic crystals  ${\rm A}_3 {\rm B}_5$ is  described
by the following Hamiltonian\cite{Dresselhaus55}:

\begin{equation}
{\cal H}_s = \gamma \sum \sigma_i k_i \left(k_{i+1}^2 - k_{i+2}^2\right),
\quad i = x, y, z, \quad i+3 \rightarrow i.
\label{HamBulk}
\end{equation}

\noindent where $\sigma_i$ are the Pauli matrices. In $[001]$  quantum
wells the size quantization gives rise to the terms in the Hamiltonian
${\cal  H}_s$,  which   are  linear  in   the  in-plane  wave   vector
$\mbox{\boldmath  $k$}  =  (k_x,  k_y)$,  in  addition  to  the  cubic
terms\cite{Altshuler81,Dyakonov86}. The corresponding Hamiltonian  for
the     conduction     band      electrons     can     be      written
as\cite{Iordanskii94,Pikus95}

\begin{equation}
{\cal H} = \frac{k^2}{2m} + (\mbox{\boldmath $\sigma$}{\bf \Omega}),
\label{HamSymm}
\end{equation}

\noindent where  $\mbox{\boldmath $\sigma$}  = (\sigma_x,  \sigma_y)$,
${\bf \Omega} = (\Omega_x, \Omega_y)$ are two-dimensional vectors with
components  in  the   plane  of  the   quantum  well.  Vector   $2{\bf
\Omega}/\hbar$ has the physical meaning of the precession vector:  its
length equals the frequency of  the spin precession and its  direction
defines the axis of the precession. The spin splitting energy is equal
to $2 \Omega$. To treat the  spin relaxation problem in the case  when
$\Omega$ is anisotropic in the $2D$ plane one has to decompose it into
orthogonal spherical harmonics:

\begin{eqnarray}
{\bf \Omega} &=& {\bf \Omega}_1 + {\bf \Omega}_3,
\nonumber \\
\Omega_{1x} &=& - \Omega_1^{(1)} \cos \varphi, \
\Omega_{3x} = - \Omega_3 \cos 3\varphi,
\nonumber \\
\Omega_{1y} &=& \ \Omega_1^{(1)} \sin \varphi, \
\Omega_{3y} = - \Omega_3 \sin 3\varphi,
\label{Omega} \\
\Omega_1^{(1)} &=& \gamma k \left( \left\langle k_z^2 \right\rangle -
{1 \over 4} k^2 \right),
\quad
\Omega_3 = \gamma {k^3 \over 4},
\nonumber
\end{eqnarray}

\noindent where $k^2 = k_x^2  + k_y^2$, $\tan \varphi =  k_x/k_y$, and
$\left\langle k_z^2 \right\rangle$ is the average squared wave  vector
in the direction  $z$, normal to  the quantum well  (in this paper  we
take $\hbar = 1$ everywhere except in final formulas).

The  spin   splitting  given   by  Eq.~(\ref{Omega})   represents  the
Dresselhaus term\cite{Dresselhaus55}. In asymmetric quantum wells  the
Hamiltonian ${\cal H}$ contain also terms of different symmetry,  i.e.
the Rashba terms\cite{Bychkov84}:

\begin{equation}
H^\prime = \alpha [\mbox{\boldmath $\sigma k$}]_z.
\label{HamAsymm}
\end{equation}

\noindent   This   term   can   be   included   in   the   Hamiltonian
Eq.~(\ref{HamSymm})  if  one  includes  additional  terms  into  ${\bf
\Omega}$:

\begin{equation}
\Omega_{1x} = \Omega_1^{(2)} \sin \varphi, \quad
\Omega_{1y} = -\Omega_1^{(2)} \cos \varphi, \quad
\Omega_1^{(2)} = \alpha k.
\label{Omega2}
\end{equation}

\noindent  In  an  uniform  electric  field  ${\cal  E}$, the constant
$\alpha$ is proportional to the field:

\begin{equation}
\alpha = \alpha_0 e {\cal E}.
\label{alphacalc}
\end{equation}

\noindent The expressions for $\alpha_0$ and $\gamma$ are given in the
Appendix. The barriers of the well give rise to another  contribution,
usually  also  linear  in  ${\cal  E}$,  which depends strongly on the
details      of      the      boundary      conditions      at     the
heterointerface\cite{Gerchikov92,PikusUnp}.

Both  terms  Eqs.~(\ref{Omega})   and  (\ref{Omega2})  give   additive
contributions  to  the  spin  relaxation  rate $1/\tau_{ij}$, which is
defined as

\begin{equation}
\frac{d s_i}{d t} = - \frac{s_j}{\tau_{ij}}.
\end{equation}

\noindent  where  $s_i$  is  an  average  projection  of  spin  on the
direction $i$. These contributions are

\begin{equation}
{1 \over \tau_{s_{xx}}} =
{1 \over 2 \tau_{s_{zz}}} =
2 \left(\Omega_1^2 \tau_1 + \Omega_3^2 \tau_3 \right),
\label{spin}
\end{equation}

\noindent where $\Omega_1^2 = {\Omega_1^{(1)}}^2 + {\Omega_1^{(2)}}^2$
and $\tau_n$, $n  = 1, 3$,  is the relaxation  time of the  respective
component of the distribution function $f_n(\mbox{\boldmath $k$}) \sim
\cos  n(\varphi_{\mbox{\boldmath  $k$}}  +  \psi_n)$  ($\psi_n$  is an
arbitrary phase):

\begin{equation}
{1 \over \tau_n} = \int W(\varphi) \left( 1 - \cos n\varphi
\right) \,d\varphi.
\label{taun}
\end{equation}

\noindent Here $W(\vartheta)$ is  the probability of scattering  by an
angle  $\vartheta$.  If  it  does  not  depend  on  $\vartheta$,   all
scattering times are equal to the elastic lifetime

\begin{equation}
\frac{1}{\tau_0} = \int W(\varphi) \,d \varphi.
\label{tau0}
\end{equation}

\noindent  When  small-angle  scattering  dominates,  $1-\cos n\varphi
\approx (n \varphi)^2/2$ and

\begin{equation}
\frac{\tau_1}{\tau_n} = n^2, \qquad (n \ge 1).
\label{tauratio}
\end{equation}

Formula Eq.~(\ref{spin})  shows that  the different  harmonics of  the
precession vector add up in  the spin relaxation rate with  the weight
equal to the  relaxation times $\tau_n$.  Unlike the spin  relaxation,
the contributions of  the different terms  in the spin  splitting into
PMC  {\em  are  not  additive}.  Furthermore, at $\Omega_1^{(1)} = \pm
\Omega_1^{(2)}$ and $\Omega_3 = 0$ the contributions of the two linear
terms Eqs.~(\ref{Omega}) and (\ref{Omega2}) exactly cancel each other,
and  the  magnetoconductivity  looks  as  if  there were no spin-orbit
interaction  at  all.  Analogous  effect  occur  in weak localization
conductance in wires\cite{Geller94}.

\subsection{Weak localization in two-dimensional structures}
\label{subsec:WeakLoc}

The weak localization contribution to the conductivity is given by the
expression\cite{Hikami80,Altshuler81}:

\begin{equation}
\Delta \sigma = - {e^2 D  \over \pi} \cdot
2 \pi \nu_0 \tau_0^2 \sum_{\alpha \beta} \int\limits_0^{q_{max}}
{{\rm  \kern.24em \vrule width.05em
height1.4ex depth-.05ex  \kern-.26em C}}
_{\alpha \beta \beta \alpha}(\mbox{\boldmath $q$}) \,
\frac{d^2 q}{(2 \pi)^2},
\end{equation}

\noindent where $\alpha$ and $\beta$ are spin indices, $q_{max}^2 = (D
\tau_1)^{-1}$, $D =  v^2 \tau_1/2$ is  the diffusion coefficient,  and
$\nu_0 =  m/2\pi$ is  the density  of states  at the  Fermi level at a
given spin projection. The matrix ${{\rm \kern.24em \vrule  width.05em
height1.4ex   depth-.05ex   \kern-.26em   C}}   _{\alpha  \beta  \beta
\alpha}(\mbox{\boldmath $q$})$  is called  Cooperon and  can be  found
from the following integral equation:

\begin{eqnarray}
&&{{\rm  \kern.24em \vrule width.05em
height1.4ex depth-.05ex  \kern-.26em C}}
_{\alpha \beta \gamma \delta}
(\mbox{\boldmath $k$}, \mbox{\boldmath $k$}^\prime, \mbox{\boldmath $q$})
=\left|V_{\mbox{\boldmath $k$},\mbox{\boldmath $k$}^\prime}\right|^2
 \delta_{\alpha \gamma} \delta_{\beta  \delta}
\label{Dyson} \\
& + & \int \frac{d^2 g}{(2 \pi)^2}
\sum_{\lambda \lambda^\prime}
  V_{\mbox{\boldmath $k$},\mbox{\boldmath $g$}}
  V_{-\mbox{\boldmath $k$},-\mbox{\boldmath $g$}}
  G^+_{\alpha \lambda}(\omega, \mbox{\boldmath $g$} + \mbox{\boldmath $q$})
  G^-_{\beta  \lambda^\prime}(\omega, -\mbox{\boldmath $g$})
\nonumber \\
& \times & {{\rm  \kern.24em \vrule width.05em
height1.4ex depth-.05ex  \kern-.26em C}}
_{\lambda \lambda^\prime \gamma \delta}
(\mbox{\boldmath $g$}, \mbox{\boldmath $k$}^\prime, \mbox{\boldmath $q$}).
\nonumber
\end{eqnarray}

\noindent Here $V_{\mbox{\boldmath $k$},\mbox{\boldmath  $k$}^\prime}$
is  a  scattering  matrix  element  (including  the  concentration  of
scatterers), which we assume here  to be diagonal in spin  indices. It
is connected with $W(\varphi)$ in Eq.~(\ref{tau0}) by the expression

\begin{equation}
W\left(\varphi_{\mbox{\boldmath $k$}} -
\varphi_{\mbox{\boldmath $k$}^\prime}\right)
= \nu_0 \left|V_{\mbox{\boldmath $k$},\mbox{\boldmath
$k$}^\prime}\right|^2,
\end{equation}

\noindent  $G^{\pm}(\omega,  \mbox{\boldmath  $k$})$  are  the Green's
functions

\begin{eqnarray}
G^{\pm}(\omega, \mbox{\boldmath $k$}) =
\left\{\omega - E(k) - (\mbox{\boldmath $\sigma$}{\bf \Omega})
\pm \frac{i}{\tau_f}\right\}^{-1}, \\
\frac{1}{\tau_f} = \frac{1}{\tau_0} + \frac{1}{\tau_\varphi}, \qquad
E(k) = \frac{k^2}{2m},
\end{eqnarray}

\noindent $\tau_\varphi$ is the  inelastic scattering time. After  the
integration by $E(g)$ in the right-hand side of Eq.~(\ref{Dyson})  the
result  is  expanded  up  to  second  order  terms  in series in small
parameters $\tau_0/\tau_\varphi$,  $\mbox{\boldmath $vq$}\tau_0$,  and
$\Omega  \tau_0$,  where  $\mbox{\boldmath  $v$} = \partial E/\partial
\mbox{\boldmath  $k$}$.  In  the  end,  the following equation for the
Cooperon is obtained:

\begin{eqnarray}
&&{{\rm  \kern.24em \vrule width.05em
height1.4ex depth-.05ex  \kern-.26em C}}_
{\mbox{\boldmath $k$}, \mbox{\boldmath $k$}^\prime}(\mbox{\boldmath
$q$})
= \left|V_{\mbox{\boldmath $k$},\mbox{\boldmath $k$}^\prime}\right|^2
+ 2 \pi \nu_0 \tau_0 \int \frac{d \varphi_g}{2 \pi}
\left|V_{\mbox{\boldmath $k$},\mbox{\boldmath $g$}}\right|^2
\nonumber \\
& & \times \Bigg\{1 - i (\mbox{\boldmath $v$}_g \mbox{\boldmath $q$}) \tau_0
- i(\mbox{\boldmath $\sigma$} + \mbox{\boldmath $\rho$}){\bf
\Omega}\tau_0
- (\mbox{\boldmath $v$}_g \mbox{\boldmath $q$})^2 \tau_0^2
\label{Cooperon} \\
& & -2(\mbox{\boldmath $\sigma$}{\bf \Omega})(
  (\mbox{\boldmath $\rho$}{\bf \Omega}) \tau_0^2
- 2 (\mbox{\boldmath $v$}_g \mbox{\boldmath $q$})
    (\mbox{\boldmath $\sigma$} + \mbox{\boldmath $\rho$}) {\bf \Omega}
\tau_0^2
- \frac{\tau_0}{\tau_\varphi} \Bigg\}
{{\rm  \kern.24em \vrule width.05em
height1.4ex depth-.05ex  \kern-.26em C}}_
{\mbox{\boldmath $g$}, \mbox{\boldmath $k$}^\prime}(\mbox{\boldmath
$q$}).
\nonumber
\end{eqnarray}

\noindent Here the Pauli  matrices \boldmath $\sigma$ \unboldmath  act
on  the  first  pair  of  spin  indices  $\alpha,  \lambda$, while the
matrices  \boldmath  $\rho$\unboldmath  --  on  the second pair $\beta
\lambda^\prime$.

The equation

\begin{equation}
{{\rm  \kern.24em \vrule width.05em
height1.4ex depth-.05ex  \kern-.26em C}}_
{\mbox{\boldmath $k$}, \mbox{\boldmath $k$}^\prime} =
\lambda \tau_0 \int W(\mbox{\boldmath $k$}^\prime, \mbox{\boldmath $g$})
{{\rm  \kern.24em \vrule width.05em
height1.4ex depth-.05ex  \kern-.26em C}}_
{\mbox{\boldmath $g$}, \mbox{\boldmath $k$}^\prime}
d \varphi_g
\label{Harmonics}
\end{equation}

\noindent has the following harmonics as its eigenfunctions:

\begin{equation}
{{\rm  \kern.24em \vrule width.05em
height1.4ex depth-.05ex  \kern-.26em C}}^n_
{\mbox{\boldmath $k$}, \mbox{\boldmath $k$}^\prime} =
{{\rm  \kern.24em \vrule width.05em
height1.4ex depth-.05ex  \kern-.26em C}}^n.
\cos n (\varphi_k - \varphi_{k^\prime} - \psi_n)
\end{equation}

\noindent  According  to  Eq.~(\ref{tau0})  the  eigenfunction  ${{\rm
\kern.24em  \vrule  width.05em  height1.4ex  depth-.05ex   \kern-.26em
C}}_0$ has  the eigenvalue  $\lambda_0=1$, while  other harmonics have
eigenvalues

\begin{equation}
\lambda_n = \left(1 -\frac{\tau_0}{\tau_n}\right)^{-1}.
\label{lambdan}
\end{equation}

\noindent   Therefore,   the   solution   of   inhomogeneous  equation
Eq.~(\ref{Cooperon}) will have large harmonic ${{\rm \kern.24em \vrule
width.05em  height1.4ex  depth-.05ex  \kern-.26em  C}}_0$,  while  the
others will  be small,  because they  appear due  to presence of small
terms   in   $q$   and   $\Omega$.   Since   the  right-hand  side  of
Eq.~(\ref{Cooperon}) contains  linear and  cubic in  $g$ terms,  in is
necessary to take  into account only  first and third  harmonics. From
Eqs.~(\ref{Cooperon}) -- (\ref{lambdan}) it follows that

\begin{eqnarray}
{{\rm  \kern.24em \vrule width.05em
height1.4ex depth-.05ex  \kern-.26em C}}^{(1)}_
{\mbox{\boldmath $g$}, \mbox{\boldmath $k$}^\prime} & = &
- i (\tau_1 - \tau_0) \left[(\mbox{\boldmath $v$}_g \mbox{\boldmath
$q$}) + (\mbox{\boldmath $\sigma$} + \mbox{\boldmath $\rho$}) {\bf
\Omega}_1(\mbox{\boldmath $g$})\right]
{{\rm  \kern.24em \vrule width.05em
height1.4ex depth-.05ex  \kern-.26em C}}^0_
{\mbox{\boldmath $g$}, \mbox{\boldmath $k$}^\prime}
\nonumber \\
{{\rm  \kern.24em \vrule width.05em
height1.4ex depth-.05ex  \kern-.26em C}}^{(3)}_
{\mbox{\boldmath $g$}, \mbox{\boldmath $k$}^\prime} & = &
- i (\tau_3 - \tau_0)
(\mbox{\boldmath $\sigma$} + \mbox{\boldmath $\rho$}) {\bf
\Omega}_3(\mbox{\boldmath $g$})
{{\rm  \kern.24em \vrule width.05em
height1.4ex depth-.05ex  \kern-.26em C}}^0_
{\mbox{\boldmath $g$}, \mbox{\boldmath $k$}^\prime}
\label{Harmon}
\end{eqnarray}

Here it  is taken  into account  that there  is a  relation similar to
Eq.~(\ref{Harmonics})      for      harmonics      $\Omega_{1\alpha}$,
$(\mbox{\boldmath      $v$}      \mbox{\boldmath      $q$})       \sim
\cos(\varphi_{\mbox{\boldmath    $g$}}    -   \varphi_{\mbox{\boldmath
$q$}})$, and $\Omega_{3\alpha}$.

Then we  substitute $  {{\rm \kern.24em  \vrule width.05em height1.4ex
depth-.05ex  \kern-.26em  C}}_  {\mbox{\boldmath $g$}, \mbox{\boldmath
$k$}^\prime}  =   {{\rm  \kern.24em   \vrule  width.05em   height1.4ex
depth-.05ex     \kern-.26em     C}}^{0}_     {\mbox{\boldmath    $g$},
\mbox{\boldmath  $k$}^\prime}  +  {{\rm  \kern.24em  \vrule width.05em
height1.4ex depth-.05ex \kern-.26em C}}^{(1)}_ {\mbox{\boldmath  $g$},
\mbox{\boldmath  $k$}^\prime}  +  {{\rm  \kern.24em  \vrule width.05em
height1.4ex depth-.05ex \kern-.26em C}}^{(3)}_ {\mbox{\boldmath  $g$},
\mbox{\boldmath  $k$}^\prime}$  into  Eq.~(\ref{Cooperon}), and, using
Eq.~(\ref{Harmon}) and retaining only the terms with zero harmonic, we
obtain   the   equation   for   ${{\rm  \kern.24em  \vrule  width.05em
height1.4ex depth-.05ex \kern-.26em C}}_{0}(\mbox{\boldmath $q$})$:

\begin{equation}
{\cal H}
{{\rm \kern.24em
            \vrule width.05em height1.4ex depth-.05ex
            \kern-.26em C}}_0
= {1 \over 2 \pi \nu_0 \tau_0^2},
\label{CooperonEq}
\end{equation}

\noindent where

\begin{eqnarray}
{\cal H} = & &{1 \over \tau_\varphi} + {1 \over 2} v^2 q^2 \tau_1 +
\nonumber \\
& & \left(\Omega_1^2\tau_1 + \Omega_3^2\tau_3\right)
\left(2+\sigma_x \rho_x + \sigma_y \rho_y \right) +
\nonumber \\
& & 2 \left(\sigma_x \rho_y + \sigma_y \rho_x\right) \Omega_1^{(1)}
\Omega_1^{(2)}\tau_1 +
\label{CoopH} \\
& &
v \tau_1 \left[ \left(\sigma_x + \rho_x\right)
\left(-\Omega_1^{(1)} q_x + \Omega_1^{(2)} q_y \right) +
\right.
\nonumber \\
& & \left.
\left(\sigma_y + \rho_y\right)
\left(\Omega_1^{(1)} q_y - \Omega_1^{(2)} q_x \right) \right].
\nonumber
\end{eqnarray}

In a magnetic field $q$ become operators with the commutator

\begin{equation}
[q_+q_-] = {\delta \over D},
\label{Commut}
\end{equation}

\noindent where $q_\pm = q_x \pm i q_y$ and

\begin{equation}
\delta = {4 e B D \over \hbar c}.
\label{delta}
\end{equation}

\noindent  This  allows  us  to  introduce  creation  and annihilation
operators $a^\dagger$ and $a$, respectively, for which $[aa^\dagger] =
1$:

\begin{equation}
D^{1/2} q_+ = \delta^{1/2} a,
\quad
D^{1/2} q_- = \delta^{1/2} a^\dagger,
\quad
D q^2 = \delta \{a a^\dagger\}.
\end{equation}

\noindent  In  the  basis  of  the  eigenfunction of the operator $\{a
a^\dagger\} = {1 \over 2} (a a^\dagger + a^\dagger a)$ these operators
have following non-zero matrix elements

\begin{eqnarray}
\left\langle n-1 \right| a \left| n \right\rangle & = &
\left\langle n \right| a^\dagger \left| n-1 \right\rangle = \sqrt{n},
\nonumber \\
\left\langle n \right| \{a a^\dagger\} \left| n \right\rangle
& = & n + {1 \over 2}.
\end{eqnarray}

In a magnetic  field, the integration  over $q$ should  be replaced by
summation over $n$. Then,

\begin{equation}
\Delta \sigma = - \frac{e^2 \delta}{4 \pi^2 \hbar}S,
\label{NMR}
\end{equation}

\noindent where

\begin{equation}
S = 2 \pi \nu_0 \tau_0^2 \sum_{\alpha, \beta, n}
{{\rm  \kern.24em \vrule width.05em
height1.4ex depth-.05ex  \kern-.26em C}}_
{\alpha \beta \beta \alpha}(n).
\label{SSum}
\end{equation}

Since Eq.~(\ref{CooperonEq}) is essentially the Green function equation
its solution can be written as:

\begin{equation}
{{\rm  \kern.24em \vrule width.05em
height1.4ex depth-.05ex  \kern-.26em C}}(n)
^{\alpha \gamma}_{\beta \delta} =
\frac{1}{2 \pi \nu_0 \tau_0^2}
\sum_{r=1}^{4} \frac{1}{E_{r, n}} \Psi_{r, n}(\alpha,\beta)
\Psi_{r, n}^*(\gamma,\delta),
\label{Psis}
\end{equation}

\noindent where $\Psi_{r,  n}$ and $E_{r,  n}$ are the  eigenfunctions
and eigenvalues of ${\cal H}$:

\begin{equation}
{\cal H}\Psi_{r, n} = E_{r, n} \Psi_{r, n}.
\end{equation}

\noindent  We  now  choose  the  basis  consisting  of  the   function
$\Psi_0(\alpha, \beta)$,  which is  antisymmetric in  spin indices and
corresponds to the total momentum $J = 0$, and of symmetric  functions
$\Psi_m$ which  correspond to  $J =  1$ and  $J_z =  m =  -1,\ 0,\ 1$.
According   to   Eq.~(\ref{Psis}),   in   this   basis   the   sum  in
Eq.~(\ref{NMR}) is

\begin{equation}
S = \sum_{n=0}^{n_{max}}
\left(- {1 \over E_{0}(n)} + \sum_{m=-1}^1 {1 \over E_{m}(n)}\right),
\label{SEr}
\end{equation}

\noindent where $n_{max} = 1/\delta\tau_1$. For the term with $J =  0$
the operator ${\cal H}$ is

\begin{equation}
{\cal H}_{0n} = \delta\{a a^\dagger\} + \frac{1}{\tau_\varphi},
\end{equation}

\noindent and, therefore,

\begin{equation}
E_0(n) = \delta \left(n + {1 \over 2} \right) +
{1 \over \tau_\varphi}.
\end{equation}

\noindent For the  term with $J  = 1$ we  can use the  relation $J_i =
(\sigma_i + \rho_i)/2$ to obtain

\begin{eqnarray}
\tilde{\cal H} = & & \delta \{a a^\dagger\} + {1 \over \tau_\varphi} +
2 \left(\Omega_1^2\tau_1 + \Omega_3^2\tau_3\right)
\left(2 - J_z^2 \right) -
\nonumber \\
& & 4 i \Omega_1^{(1)} \Omega_1^{(2)}\tau_1
\left( J_+^2 - J_-^2 \right) + 2(\delta \tau_1)^{1/2}
\label{ErMag} \\
& &
\times \left[-\Omega_1^{(1)} \left(J_+ a + J_- a^\dagger \right) +
      i\Omega_1^{(2)} \left(J_+ a^\dagger - J_- a \right) \right].
\nonumber
\end{eqnarray}

\noindent where $J_\pm = (J_x \pm iJ_y)/\sqrt{2}$.

When  $\Omega_1^{(2)}  =  0$  (or  $\Omega_1^{(1)} = 0$), the operator
Eq.~(\ref{ErMag})  can  be  reduced  to  a  block-diagonal  form  with
$3\times3$ blocks if one uses the basis of functions $\Psi_n =  {\bf(}
f_1(n) F_{n-1}, f_0(n) F_n, f_{-1}(n) F_{n+1}{\bf)}$, where $F_n$  are
the eigenfunctions of the operator $\{a a^\dagger\}$ and $f_m$ are the
eigenfunctions of $J_z$ (for $\Omega_1^{(1)} = 0$ the basis is $\Psi_n
= {\bf(} f_1(n) F_{n+1}, f_0(n) F_n, f_{-1}(n) F_{n-1}{\bf)}$).  Using
the formula

\begin{equation}
\sum_m \frac{1}{E_m} = \sum_m\frac{\left|D_{mm}\right|}{|D|},
\label{Minors}
\end{equation}

\noindent   where   $|D|$   is   the   determinant   of   ${\cal   H}$
{\bf(}Eq.~(\ref{ErMag}){\bf)} and $\left|D_{mm}\right|$ are its minors
of  diagonal  elements  $D_{mm}$,  the  sum  in Eq.~(\ref{SEr}) can be
immediately      calculated\cite{Iordanskii94}.      According      to
Eqs.~(\ref{NMR}), (\ref{ErMag}), and (\ref{Minors}),

\begin{eqnarray}
\Delta&&\sigma(B) = - \frac{e^2}{4 \pi^2 \hbar}
\Biggl\{
\frac{1}{a_0} +
\frac{2a_0 + 1 + \frac{H_{SO}}{B}}
{a_1 \left(a_0 + \frac{H_{SO}}{B}\right) - 2 \frac{H^\prime_{SO}}{B}}
\nonumber \\
& & - \sum_{n=0}^{\infty}
\Biggl(\frac{3}{n}
\label{NMRIPL} \\
& & -\frac{3a_n^2 + 2 a_n\frac{H_{SO}}{B}-1-2(2n+1)\frac{H^\prime_{SO}}{B}}
{\left(a_n + \frac{H_{SO}}{B}\right)a_{n-1}a_{n+1} -
2 \frac{H^\prime_{SO}}{B}\left[(2n+1)a_n - 1\right]}
\Biggr)
\nonumber \\
& &
+ 2 \ln \frac{H_{tr}}{B} + \Psi\left(\frac{1}{2} + \frac{H_\varphi}{B}
\right) + 3 C
\Biggr\},
\nonumber
\end{eqnarray}

\noindent where $C$ is the Euler's constant,

\begin{eqnarray}
a_n & = & n + \frac{1}{2} + \frac{H_\varphi}{B} + \frac{H_{SO}}{B},
\nonumber \\
H_\varphi & = & {c \over 4 e \hbar D \tau_\varphi}, \
{B \over H_\varphi} = \delta\tau_\varphi, \
H_{tr} = {c \over 4 e \hbar D \tau_1},
\label{Fields} \\
H_{SO} & = & {c \over 4 \hbar e D} \left(2\Omega_1^2\tau_1 +
2\Omega_3^2 \tau_3 \right), \
H^\prime_{SO}   =  H^{(1)}_{SO} {\rm \ or \ } H^{(2)}_{SO},
\nonumber \\
H^{(1)}_{SO} & = & {c \over 4 \hbar e D} {2\Omega_1^{(1)}}^2\tau_1, \
H^{(2)}_{SO}   =   {c \over 4 \hbar e D} {2\Omega_1^{(2)}}^2\tau_1,
\nonumber
\end{eqnarray}

\noindent and $\Psi$ is a digamma-function.

If both $\Omega_1^{(1)}  = \Omega_1^{(2)} =  0$ and only  the cubic in
$k$ term with $\Omega_3$ is present, the expression Eq.~(\ref{NMRIPL})
can be further reduced to  the formula, which was obtained  earlier in
Ref.~\onlinecite{Altshuler81}:

\begin{eqnarray}
\Delta && \sigma(B) - \Delta \sigma(0) =  \frac{e^2}{2 \pi^2 \hbar}
\Biggl\{
\Psi\left(\frac{1}{2} + \frac{H_\varphi}{B} + \frac{H_{SO}}{B}\right) +
\nonumber \\
& & \frac{1}{2}
\Psi\left(\frac{1}{2} + \frac{H_\varphi}{B}+2\frac{H_{SO}}{B}\right)-
\frac{1}{2}
\Psi\left(\frac{1}{2} + \frac{H_\varphi}{B}\right)-
\label{NMRHLN} \\
& & \ln \frac{H_\varphi + H_{SO}}{B} -
\frac{1}{2} \ln \frac{H_\varphi + 2H_{SO}}{B} +
\frac{1}{2} \ln \frac{H_\varphi}{B}
\Biggr\}.
\nonumber
\end{eqnarray}

\noindent Note that, according to Ref.~\onlinecite{Iordanskii94},  the
value of $H_{SO}$ is twice that used in Ref.~\onlinecite{Altshuler81}.

The case when $\Omega_1^{(1)} = \pm \Omega_1^{(2)}$ and $\Omega_3 = 0$
is  a  special  one.  In  this  case the operator Eq.~(\ref{CoopH}) is
diagonal in the  basis of functions  $\Psi_m$ if one  uses coordinates
$x^\prime \parallel (110)$ and $y^\prime \parallel (1\bar{1}0)$:

\begin{equation}
{\cal H}_{mm^\prime} = \left\{\frac{1}{\tau_\varphi} + D\left[
q_{x^\prime}^2 + \left(q_{y^\prime} + q_{y^\prime m}^0\right)^2
\right]\right\}\delta_{mm^\prime},
\end{equation}

\noindent where $q_{y^\prime m}^0 = 2 \Omega_1\sqrt{\tau_1/D}m$. Since
the  commutation  relations  Eq.~(\ref{Commut})  do  not  change  when
$q_{y^\prime}$ is shifted by $q_{y^\prime}^0$, the spin splitting does
not manifest itself in the magnetoconductivity, which is given by  the
simple formula\cite{Hikami80}

\begin{equation}
\Delta \sigma(B) - \Delta \sigma(0) = \frac{e^2}{2 \pi^2 \hbar}
\left\{\Psi\left(\frac{1}{2} + \frac{H_\varphi}{B}\right)-
\ln \frac{H_\varphi}{B}\right\}.
\end{equation}

\noindent It was  demonstrated in Ref.~\onlinecite{Pikus95}  that this
result appears because, when $\Omega_1^{(1)} = \pm \Omega_1^{(2)}$ and
$\Omega_3  =  0$,  the  total  spin  rotation for the motion along any
closed trajectory is exactly zero.

When $\Omega_1^{(1)}$ and $\Omega_1^{(2)}$ are not equal or  $\Omega_3
\ne 0$, the  only way to  find eigenvalues $E_{mn}$  is to diagonalize
numerically the matrix  $\tilde{\cal H}$. The  number of elements  one
has to take for a given  value of magnetic field $B$, or  $\delta$, is
at least  $n_{max}= 1/\delta\tau_1$  and increases  infinitely as  $B$
approaches $0$. Note that the  size of the matrix $\tilde{\cal  H}$ is
$N  =  3  n_{max}$.  For  the  detail  of the numerical procedure, see
Ref.~\onlinecite{Pikus95}.

\subsection{Elliott-Yafet spin relaxation mechanism}
\label{subsec:EY}

It follows from Ref.~\onlinecite{Hikami80} that in order to take  into
account the Elliott-Yafet spin relaxation  mechanism one has to add  a
new term to the Hamiltonian Eq.~(\ref{ErMag}):

\begin{equation}
{\cal H}_{EY} = \frac{1}{\tau_{s_{EY}}}J_z^2,
\label{HamEY}
\end{equation}

\noindent where, according to Ref.~\onlinecite{Pikus84},

\begin{eqnarray}
\frac{1}{\tau_{s_{EY}}} & = & \frac{1}{\tau_2}\left(\kappa^2
\beta\right)^2,
\label{tauEY} \\
\beta & = & \frac{\hbar^2}{3 m} \frac{\Delta \left(E_g - \frac{\Delta}{2}
\right)}
{E_g^2 \left(E_g - \frac{\Delta}{3}\right)},
\end{eqnarray}

\noindent and $\tau_2$ is defined by Eq.~(\ref{taun}).

As  a  result,  in  the  first  and  fourth  terms  of the formula for
magnetoconductivity Eq.~(\ref{NMRHLN}) $H_{SO}$ should be replaced  by
$H_{SO} + H_{EY}$, where $H_{EY}$ is

\begin{equation}
H_{EY} = \frac{c}{4 \hbar e D \tau_{s_{EY}}}.
\end{equation}

\noindent It follows from Eq.~(\ref{tauEY}) that

\begin{equation}
\frac{H_{EY}}{H_{tr}} = \left(2 \pi N_s \beta\right)^2 \,
\frac{\tau_1}{\tau_2}.
\label{EYField}
\end{equation}

\section{Experimental Procedures}
\label{sec:Experiment}

\subsection{Samples}
\label{subsec:samples}


\begin{figure}[t]

\epsfxsize=3 in
\epsffile{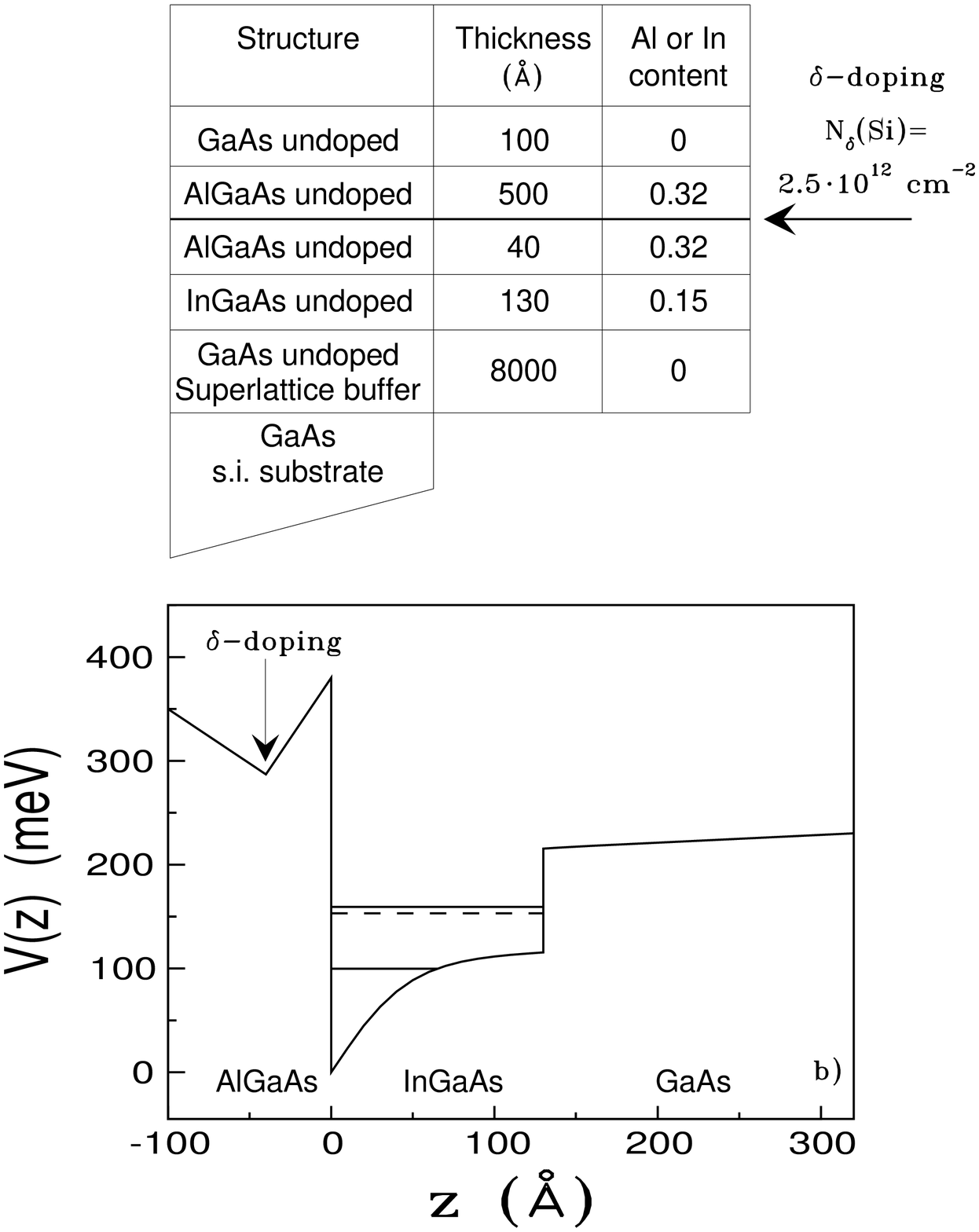}

\caption{
Sample structure  (a) and  band diagram  (b) for  GaInAs quantum  well
(sample {\sl B1}) as  obtained from self-consistent calculations.  The
first two energy levels  in the well are  shown by solid lines,  Fermi
energy is shown by a dotted line.
\label{SamplePlot}}
\end{figure}



\begin{figure}[t]

\epsfxsize=3 in
\epsffile{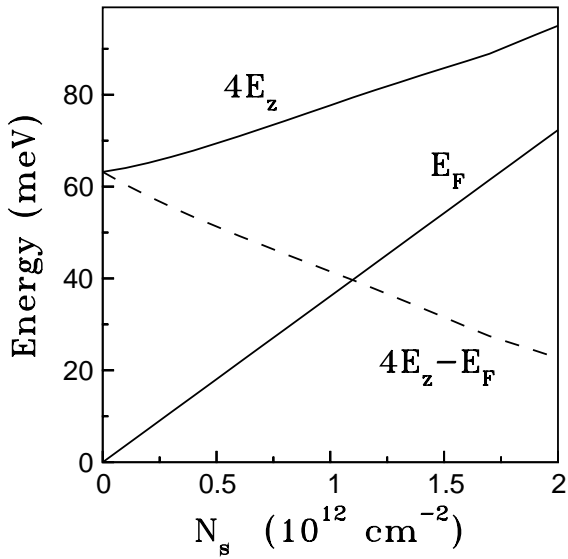}

\caption{
Energies determining the Dresselhaus  spin splitting as a  function of
electron density $N_s$. Fermi energy $E_F$ and quadrupled mean kinetic
energy $4E_Z$ of the motion in the growth direction are shown by solid
lines.  Dotted  line  shows  the  difference  $4E_Z - E_F$ that enters
Eq.~(\protect\ref{Fields}) for $H^{(1)}_{SO}$.
\label{EFermiPlot}}
\end{figure}



\narrowtext
\begin{table}[b]
\caption{Sample parameters:  electron density  $N_s$, mobility  $\mu$,
transport  magnetic  field  $H_{tr}$  (Eq.~(\protect\ref{Fields}), and
momentum relaxation time $\tau_1$.
\label{SampleTable}}
\begin{tabular}{cccccc}
$N_s {\ \rm (10^{12}\ cm^{-2})}$ & $\mu {\ \rm (m^2/V s)}$ &
$H_{tr}$ (Gs) & $\tau_1$ (ps) & sample & spacer \\
\tableline
0.98 & 2.96 & 14 & 1.2 & {\sl A1} & 6  nm \\   
1.1 & 3.72 & 7.9 & 1.5 & {\sl A2} & 6  nm \\   
1.15 & 4.11 & 6.2 & 1.7 & {\sl A3}& 6  nm \\   
1.34 & 1.94 & 24 & 0.8 & {\sl B1} & 4  nm \\   
1.61 & 1.85 & 22 & 0.8 & {\sl C1} & 2  nm \\   
1.76 & 1.63 & 26 & 0.7 & {\sl C2} & 2  nm \\   
1.79 & 1.57 & 27 & 0.7 & {\sl C3} & 2  nm \\   
1.85 & 1.43 & 32 & 0.6 & {\sl C4} & 2  nm \\   
\end{tabular}
\end{table}
\narrowtext


Three  AlGaAs/InGaAs/GaAs  pseudomorphic  quantum  wells were studied.
They were  grown by  the molecular  beam epitaxy  technique. The layer
sequence of the structure was of  the standard HEMT type and is  shown
in Fig.~\ref{SamplePlot}. The two-dimensional electron gas was  formed
in the 13 nm thick  InGaAs layer. Samples were $\delta$-doped  with Si
(doping density  $N_d=2.5 \cdot  10^{12}\ {\rm  cm}^{-2})$. Samples of
the type {\sl A} had a spacer  thickness of 6 nm, samples of the  type
{\sl B} had a 4  nm spacer and samples of  the type {\sl C} had  a 2nm
spacer. The samples had the Hall  bar geometry with the length of  1.0
mm and the width of 0.1  mm with two current and four  voltage probes.
The  distance  between  voltage  probes  was  0.3 mm. The samples were
independently characterized by luminescence, high field transport, and
cyclotron  emission  experiments\cite{Litwin94}.  The  parameters  are
listed in Table~\ref{SampleTable}. In  order to study the  behavior of
the structures as a function of electron density $N_s$, the metastable
properties of the DX-Si centers present in AlGaAs layer were employed.
Different concentrations  were obtained  by cooling  sample slowly  in
dark and then by illuminating it gradually by a light-emitting  diode.
This allowed us to tune carrier density from $0.98 \cdot 10^{12}\ {\rm
cm}^{-2}$ to $1.95 \cdot 10^{12}\ {\rm cm}^{-2}$. We have measured the
Hall effect and Shubnikov-de-Haas oscillations 
to determine  $N_s$ and to verify  that
in all samples only the  lowest subband is occupied. To  calculate the
energy levels in investigated quantum wells we first self-consistently
calculate the $2D$ wavefunctions, using the envelope function approach
in  the  Hartree  approximation\cite{Mosser72,Mosser82}. The potential
entering into the zero-magnetic  field Hamiltonian takes into  account
the  conduction  band  offset  at  each  interface, and includes, in a
self-consistent way, the electrostatic potential curvature due to  the
finite extent of the electron wavefunction. The boundary condition for
the integration  of Poisson  equation within  the $2D$  channel is the
value  of  the  built-in  electric  field  in  the buffer layer on the
substrate side of the $2D$ channel. It originates from the pinning  of
the  Fermi  level  near  midgap  in semiinsulating GaAs substrate. Any
nonparabolicity effects  on the  effective masses  were neglected. The
calculations  were  performed  for  the  temperature 4.2 K. Results of
calculations  are  shown  in  Fig.~\ref{SamplePlot}.  With  increasing
concentration both Fermi  energy and kinetic  energy of the  motion in
the growth direction increase. Their exact concentration  dependencies
should be determined to  calculate spin splitting and  spin relaxation
times. For every  carrier density $N_s$  the expectation value  of the
$z$-component    of    the    kinetic    energy    was     calculated.
Figure~\ref{EFermiPlot} shows the result of such calculations for  the
quantum wells used in our  experiments. We also show the  Fermi energy
as a function of carrier density $N_s$.

\subsection{Magnetic field generation and stability}
\label{subsec:magnet}

We have used a system  of two superconducting coils (8T/8T)  placed in
the same  cryostat. This  system was  earlier used  to study cyclotron
emission from the same  samples {\sl A}, {\sl  B}, and {\sl C}  and to
determine their effective  masses\cite{Knap92}. The sample  was placed
in the center of the first coil. To generate the stable weak  magnetic
field,  necessary  for  the  antilocalization  measurements  we used a
spread field of the second coil  to compensate the field in the  first
one.  The  magnetic  field  scale  was  determined  on  the  basis  of
measurements of the Hall voltages induced on the sample by both coils.
Typically the constant  magnetic field in  the sample coil  was of the
order of 400 Gauss  and it was compensated  by tuning the second  coil
field in the  range from 12  to 14 kGauss.  This way, both  coils were
operated in  a stable  and reproducible  manner giving  in the  sample
space  magnetic  fields  from  -  30  Gauss to +30 Gauss. Small sample
dimensions  and  the  geometry  of  the  coils  gave  good magnetic   field
uniformity. We estimate  thet the magnetic field have varied by  less than  0.1 Gauss  over the
sample.

\subsection{Conductivity  measurements  and  temperature  control}
\label{subsec:temp}


\begin{figure}[t]

\epsfxsize=3 in
\epsffile{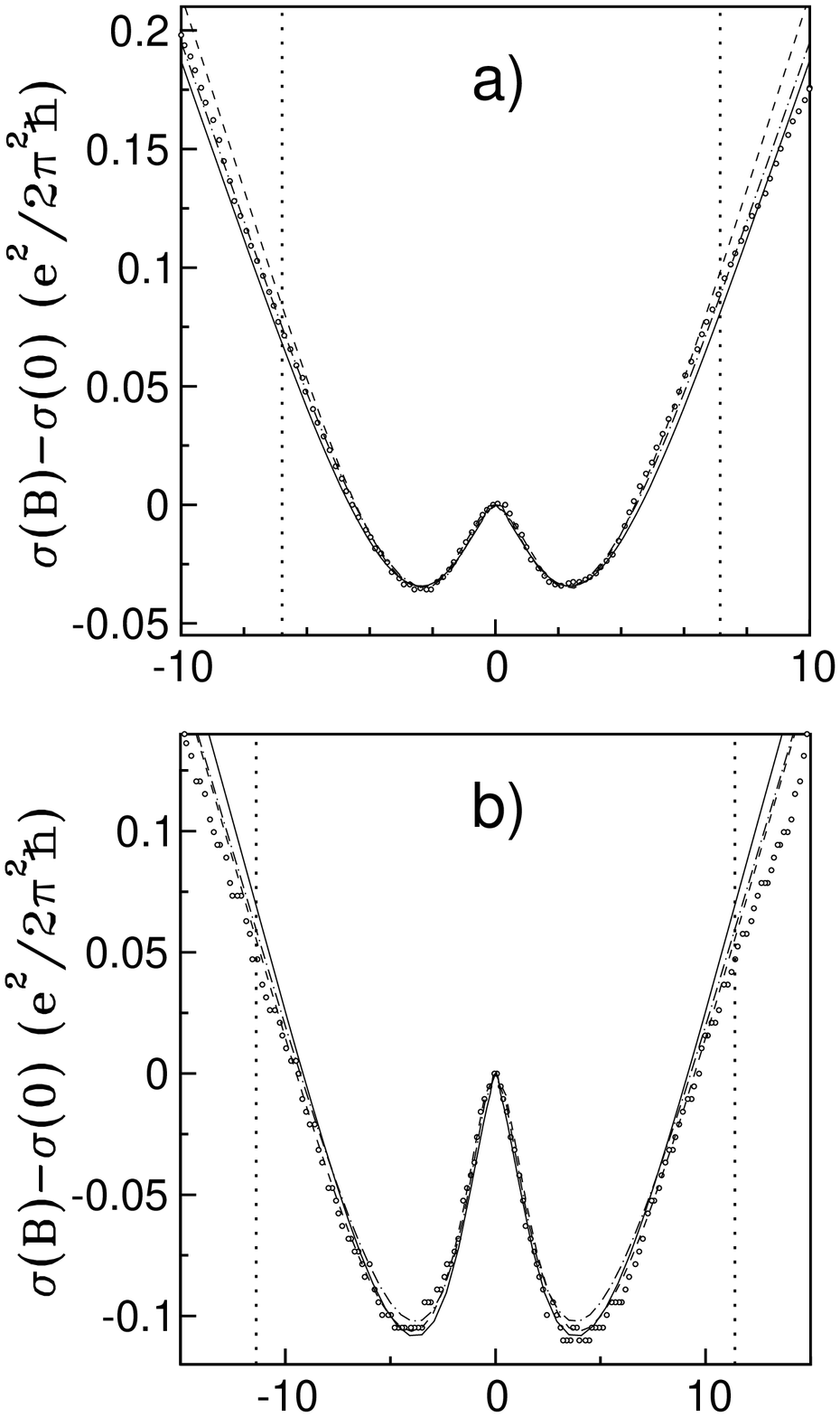}

\end{figure}


We have used  the standard direct  current (DC) method  to measure the
conductivity with currents  less than 20  microampers to avoid  sample
heating. A high precision voltmeter capable of measuring nV changes on
mV signals was used to measure the conductivity and Hall voltages. The
whole  system  was  computer  controlled.  To  avoid  mechanical   and
temperature instabilities, the sample was not directly immersed in the
liquid helium but was enclosed  in the vacuum tight sample  holder and
cooled by helium exchange gas under the 50 mbar pressure. A calibrated
Allan--Bradley resistor placed near the sample was used to measure the
temperature  which   was  stabilized   between  4.2K   and  4.3K.  The
experimental    arrangement    allowed    simultaneous   complementary
Shubnikov-de Haas  and Hall  effect measurements  to determine carrier
mobility   and   concentration   for   different  sample  illumination
intensities.

\section{Results and Discussion}
\label{sec:Results}

\subsection{General comments}
\label{subsec:general}

For all samples and for all carrier densities, the magnetoconductivity
was  a  non-monotonic  function  of  the  magnetic  field.  As we have
mentioned before, presence of a minimum on the $\sigma(B)$ curves is a
definite  sign  that  the  dominant  spin-relaxation  mechanism is the
Dyakonov--Perel  one.   For  the   Elliott-Yafet  mechanism   in  $2D$
structures, the  contribution of  the singlet  state with  $J =  0$ is
exactly canceled by one of the triplet states, namely the one with $J
=   1$   and   $J_z   =   0$,   which   is  immediately  evident  from
Eq.~(\ref{HamEY}). Using  Eq.~(\ref{EYField}) one  can show  that even
for the highest density  $N_s = 2 \cdot  10^{12} {\rm \ cm^{-2}}$  and
$\tau_1/\tau_2 = 4$, the  characteristic magnetic field $H_{EY}$  does
not  exceed  $4\cdot  10^{-4}\,H_{tr}$,  which  is  much  smaller than
$H_{SO}$. For the scattering on paramagnetic impurities, the  negative
magnetoconductivity at lowest fields does  not exist both in $2D$  and
$3D$ systems.

As we have already noted, the theory presented in this paper uses  the
diffusion approximation, which  is valid only  when all of  the fields
$H_\varphi$ and  $H_{SO}$ are  smaller than  $H_{tr}$. Kinetic theory,
which   is   free   from    this   limitation,   was   developed    in
Refs.~\onlinecite{Chak86,Kawabata94,Dyakonov94,Zduniak95} for the case
of isotropic  scattering and  with spin  relaxation considered  in the
framework of AALKh  theory. The comparison  with the diffusion  theory
shows that in magnetic field $B = 0.4 H_{tr}$ the latter has an  error
of 6\%\cite{Zduniak95}. For the purpose of comparison with theory,  we
have selected only samples with $B$ at the minimum of $\sigma$ smaller
than $0.4 H_{tr}$.

\subsection{Description of fitting procedure}
\label{subsec:Fit}


\begin{figure}[t]

\epsfxsize=3 in
\epsffile{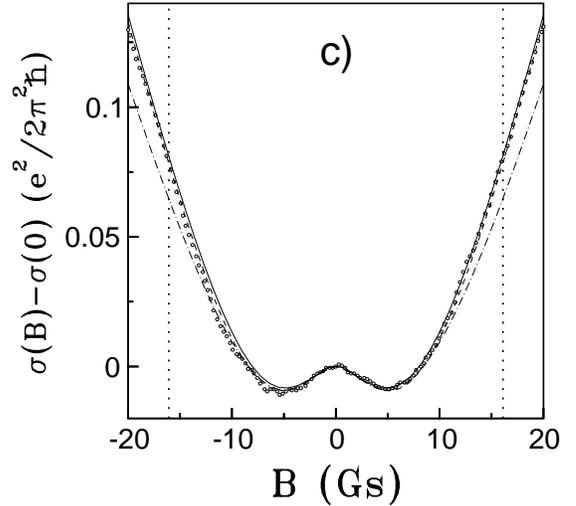}

\caption{
Experimental   results   (circles)   and   theoretical  fits  for  the
magnetoconductivity  $\sigma(B)  -  \sigma(0)$  for  three   different
samples: a) --  {\sl A1}, b)  -- {\sl B1},  and c) --  {\sl C4}. Solid
lines     show     results     of     the     theory    outlined    in
Sec.~\protect\ref{sec:Theory}.     Best     fits     obtained     from
Eqs.~(\protect\ref{NMRIPL})  and  (\protect\ref{NMRHLN})  are shown by
dashed and  dotted-dashed lines,  respectively. Dotted  vertical lines
show  the  values  $B  =  0.5  H_{tr}$,  which  limit the intervals of
applicability   of   all   three   theories.  The  fitting  parameters
$H^{(1)}_{SO}$, $H^{(2)}_{SO}$, $H_{SO}$, and $H_\varphi$ are given in
Table~\protect\ref{FieldsTable}.
\label{NMRPlot}}
\end{figure}


The experimental data for each  sample are fitted with the  results of
three  different  theoretical  models.  First  one is the AALKh theory
\cite{Altshuler81} Eq.~(\ref{NMRHLN}) and has $H_{SO}$ and $H_\varphi$
as  fitting  parameters.  The  second  one corresponds to the physical
situation   where   one   of   the   linear  terms  $H^{(1)}_{SO}$  or
$H^{(2)}_{SO}$ dominates, and the Eq.~(\ref{NMRIPL}) can be used  with
the    fitting     parameters    $H_{SO}$,     $H_{SO}^\prime$,    and
$H_\varphi$\cite{Iordanskii94}. The last theory takes into account all
the terms  $H_{SO}$, $H^{(1)}_{SO}$,  and $H^{(2)}_{SO}$  exactly. The
results of this theory  were obtained by numerical  diagonalization of
the      matrix      Eq.~(\ref{ErMag}),      as      described      in
Sec.~\ref{subsec:WeakLoc} and  Ref.~\onlinecite{Pikus95}. The  fitting
parameters in this case are $H_{SO}$, $H^{(1)}_{SO}$,  $H^{(2)}_{SO}$,
and $H_\varphi$ (see Table~\ref{FieldsTable}).

The  fitting  of  the  experimental  data  by  Eqs.~(\ref{NMRIPL}) and
(\ref{NMRHLN})  was  done  by  weighted  explicit  orthogonal distance
regression using the software package ODRPACK \cite{ODR}. The  weights
were selected to increase the importance of the low-field part of  the
magnetoconductivity curve. The calculation of the  magnetoconductivity
by  numerical  diagonalization  of  the  matrix  Eq.~(\ref{ErMag}), as
described  in  Sec.~\ref{subsec:WeakLoc},  requires  large  amounts of
computer time, and  we could not  afford to use  the automated fitting
with  these  results.  The   fitting  was  done  ``by   hand'',  using
empirically  gained  knowledge  on  how  changing  different   fitting
parameters affect the magnetoconductivity curve.

\subsection{Experimental results}
\label{subsec:exp}


\mediumtext
\begin{table*}[t]
\caption{Parameters of the best fits for three samples {\sl A1},  {\sl
B1}, and {\sl C4} (shown in Fig.~\protect\ref{NMRPlot} a), b), and c),
respectively)     as     obtained      from     the     theory      of
Sec.~\protect\ref{sec:Theory}   (rows   I),   from   the   theory   of
Ref.~\protect\onlinecite{Iordanskii94}  and   Eq.~\protect\ref{NMRIPL}
(rows       II),       and       from       the       theory        of
Ref.~\protect\onlinecite{Altshuler81}   and   Eq.~\protect\ref{NMRHLN}
(rows III). All magnetic fields are in Gauss.
\label{FieldsTable}}
\begin{tabular}{ccccccc}
Sample & Theory &
$H^{(1)}_{SO}$ & $H^{(2)}_{SO}$ & $H_{SO}$ & $H_\varphi$ &
$H^{(3)}_{SO} = H_{SO} - H^{(1)}_{SO} - H^{(2)}_{SO}$ \\
\tableline
         &  I  &  0.62 & 1.41 & 2.69 & 0.66 & 0.66 \\
{\sl A1} &  II &  0    & 0.03 & 0.85 & 0.66 & 0.82 \\
         & III &  0    & 0    & 0.77 & 0.59 & 0.77 \\
\tableline
         &  I  &  0.66 & 1.91 & 3.52 & 0.60 & 0.96 \\
{\sl B1} &  II &  0    & 0.87 & 1.89 & 0.58 & 1.02 \\
         & III &  0    & 0    & 1.08 & 0.53 & 1.08 \\
\tableline
         &  I  &  0.34 & 4.32 & 5.98 & 3.03 & 1.33 \\
{\sl C4} &  II &  0    & 3.97 & 5.30 & 3.03 & 1.51 \\
         & III &  0    & 0    & 2.18 & 2.38 & 2.18 \\
\end{tabular}
\end{table*}
\narrowtext


In Fig.~\ref{NMRPlot} a--c we show the results of the measurements  of
the conductivity $\sigma$  as a function  of magnetic field  for three
different  samples.  To  compare  the  results  for  different carrier
densities we plot $\sigma(B)-\sigma(0)$ in units of $e^2/2 \pi^2 \hbar
= 1.23 10^{-5}\ \Omega^{-1}$. The value of $\sigma(B)-\sigma(0)$ gives
the conductivity change induced by the applied magnetic field and  can
be directly compared  with theory. The  circles show the  experimental
data, the results of the theory presented in Sec.~\ref{sec:Theory} are
shown   by   solid   lines.   The   values   of  parameters  $H_{SO}$,
$H^{(1)}_{SO}$, and  $H^{(2)}_{SO}$, as  well as  values of  $N_s$ and
$H_{tr}$, are given in Table~\ref{FieldsTable}.

Before the quantitative  analysis of the  experimental data, we  would
like to point out some of their general features. The position of  the
characteristic  conductivity  minimum  which  shifts  from  2.5  Gs in
Fig.~\ref{NMRPlot}a  to  5   Gs  in  Fig.~\ref{NMRPlot}c   is  largely
determined  by  the  value  of  $H_{SO}$,  and,  hence,  by  the  spin
relaxation rate.  With increasing  carrier density  $N_s$ this minimum
shifts towards higher magnetic  fields. This indicates an  increase in
the efficiency of the spin  relaxation. One can also observe  that the
minimum  becomes  more  pronounced  when  the  ratio $H_{SO}/H\varphi$
increases: the minimal value of $\sigma(B) - \sigma(0)$ is about $0.04
e^2/2\pi^2\hbar$ for the sample  {\sl A1}, $0.01 e^2/2\pi^2\hbar$  for
the sample {\sl C4}, but increases to $0.11 e^2/2\pi^2\hbar$ for  {\sl
B1}.  This  shows  that  the  magnitude of the antilocalization effect
depends  strongly  on  the  ratio  of  the  phase-breaking  and   spin
relaxation rates. Small phase-breaking  rate and fast spin  relaxation
increase the  magnitude of  the antilocalization  phenomenon. When the
two rates are comparable, the antilocalization minimum almost vanishes
(this can be seen in Fig.~\ref{NMRPlot}c for the sample {\sl C4}).

In Fig.~\ref{NMRPlot}a for the sample  {\sl A1} the dashed line  shows
the   best   fit   obtained   using   Eq.~(\ref{NMRIPL}),   i.e.  with
$H^{(2)}_{SO}  =  0$.   The  best  fit   value  of  $H^\prime_{SO}   =
H^{(1)}_{SO} = 0.03 {\rm  Gs}$ is also close  to 0. Hence, the  dashed
curve almost coincides  with the dashed-dotted  line, which shows  the
result  of  AALKh  theory,  Eq.~(\ref{NMRHLN}).  Both theories fit the
experimental  data  seemingly  quite  well.  However,  the  values  of
parameters required  to achieve  this agreement  ($H_{SO} \approx 0.8$
and  $H^{(2)}_{SO}  \approx  0$)  are  in  a  sharp contradiction with
theoretical calculations of $H_{SO}$ and experimental measurements  of
$\gamma$, while the theory presented in this paper fits the experiment
using  the  parameters  $\alpha$  and  $\gamma$ which agree with other
measurements  and  calculations  (see  Sec.~\ref{subsec:density},  the
Appendix,                                                          and
Refs.~\onlinecite{Pikus84,Vogl83,Kobayashi82,Cardona94,Cardona95}).

The results for the sample {\sl B1} are shown in Fig.~\ref{NMRPlot} b.
Again, the  dashed line  shows the  fit by  Eq.~(\ref{NMRIPL}) and the
dashed-dotted line -- by Eq.~(\ref{NMRHLN}). One can see that in  this
case the theory with both $H^{(1)}_{SO}$ and $H^{(2)}_{SO}$, presented
in this paper (solid line),  gives somewhat better agreement with  the
experiment in the  vicinity of the  conductivity minimum. The  general
agreement of  all curves  with experiment  is of  similar quality, but
again  in  order  to  bring  Eqs.~(\ref{NMRIPL}) and (\ref{NMRHLN}) in
agreement  with  experiment  one  has  to  use  unrealistic  values of
$H_{SO}$ and $H^\prime_{SO}$.

Fig.~\ref{NMRPlot} c shows  the results for  the sample {\sl  C4}. The
dotted-dashed line in Fig.~\ref{NMRPlot}  c shows the result  of AALKh
theory, Eq.~(\ref{NMRHLN}). One can see that for $B \ge 10 {\ \rm Gs}$
this curve deviates from the experimental results quite significantly.
For this sample, as  well as for two  other samples {\sl C2}  and {\sl
C3} with large electron densities and $H^{(2)}_{SO} \gg H^{(1)}_{SO}$,
we have taken $H^{(1)}_{SO}$ to be equal to its theoretical value  for
$\gamma = 24 {\ \rm eV\,\AA^3}$. One can see from Fig.~\ref{NMRPlot} c
that the solid curve, computed for $H^{(1)}_{SO}=0.34 {\ \rm Gs}$  and
$H^{(2)}_{SO}  =  4.32  {\  \rm  Gs}$,  practically coincides with the
curve,  computed  using  Eq.~(\ref{NMRIPL})  for  $H^{(1)}_{SO}=0$ and
$H^{(2)}_{SO} =3.97 {\ \rm Gs}$.  This means that for large  $N_s$ the
experiment  allows  to  measure  only  the  difference $H^{(2)}_{SO} -
H^{(1)}_{SO}$. The  discussion above  shows that  the new  theoretical
approaches developed in this work allow to improve the description  of
the  magnetoconductivity   dependencies  and   to  obtain   meaningful
parameters from the fits. In the next section we show that using the
compkete theoretical description with $H^{(1)}_{SO}$, $H^{(2)}_{SO}$,
and $H_{SO}$ as the parameters one can get a consistent description of
experimental data for samples with different carrier densities.

\subsection{Carrier density dependencies}
\label{subsec:density}


\begin{figure}[t]

\epsfxsize=3 in
\epsffile{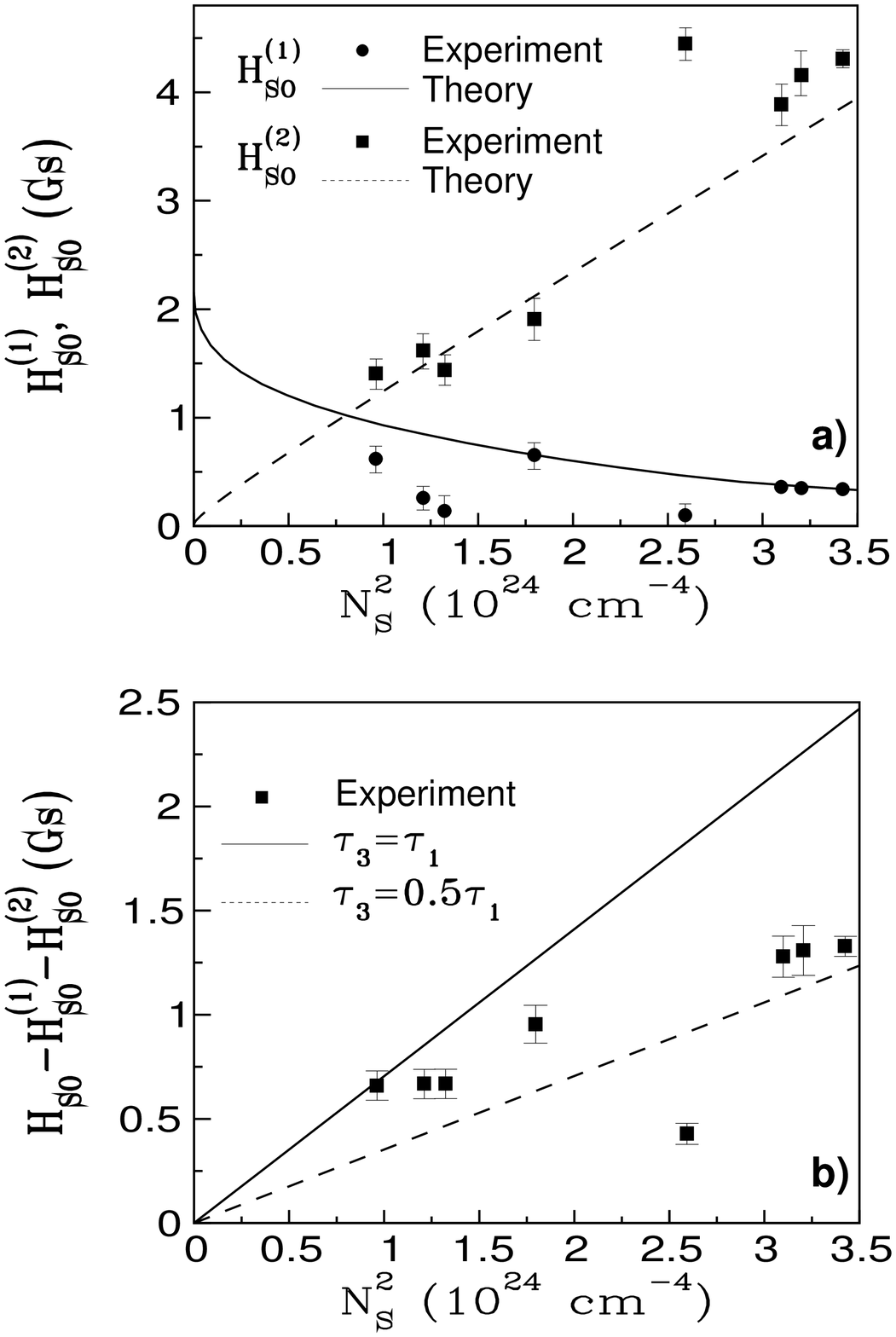}

\caption{
Characteristic magnetic fields as  a function of the  electron density
$N_s$. \protect\\
a) --  Density dependencies  of the  Dresselhaus ($H^{(1)}_{SO}$)  and
Rashba ($H^{(2)}_{SO}$)  linear terms  are shown  by dotted  and solid
lines,   respectively.   Calculations    were   done   according    to
Eq.~\protect\ref{Fields}  with  $\gamma  =  24  {\  \rm eV \AA^3}$ and
$\alpha_0 = 7.3 {\ \rm \AA^2}$. The values of these fields as obtained
from best fit with the Sec.~\protect\ref{sec:Theory} theory are  shown
by squares (Rashba term) and circles (Dresselhaus term). \protect\\
b)  --   The  Dresselhaus   cubic  term   $H_{SO}  -   H^{(1)}_{SO}  -
H^{(2)}_{SO}$ as a function of  $N_s$. The lines are calculated  using
Eq.~\protect\ref{Fields} for  $\gamma =  24 {\  \rm eV  \AA^3}$. Solid
line shows results for  an isotropic scattering, $\tau_1/\tau_3  = 1$,
Dotted line -- for $\tau_1/\tau_3 = 2$.
\label{HSOPlot}}
\end{figure}


In  Fig.~\ref{HSOPlot}  we  show  the  values  of  $H^{(1)}_{SO}$  and
$H^{(2)}_{SO}$  as  a  function  of  $N_s^2$  for  all samples we have
studied, as obtained from the  fitting of the experimental results  by
our theory.  We also  show the  theoretical curves  for these  fields,
calculated      using      Eqs.~(\ref{Omega}--\ref{alphacalc})     and
(\ref{Fields}):

\begin{eqnarray}
H^{(1)}_{SO} = \eta_1 \gamma^2 N_s^2 \left(\frac{m}{m_0}\right)^2
\left(4\frac{E_Z}{E_F} - 1\right)^2,
\nonumber \\
H^{(2)}_{SO} = \eta_2 \alpha_0^2 N_s^2 \left(\frac{m}{m_0}\right)^2
\frac{1}{\kappa^2} \left(2\frac{N_0}{N_s} + 1\right)^2,
\label{Hcurves}
\end{eqnarray}

\noindent where $N_0$  is the charge  density in the  depletion layer,
$E_z =  \hbar^2 \langle  k_z^2 \rangle/2  m$ is  the kinetic energy of
motion in $z$-direction, and

\begin{equation}
\eta_1 = \frac{\pi^2 c m_0^2}{4 e \hbar^3},\
\eta_2 = \frac{4\pi^2 c m_0^2 e^3}{\hbar^3}.
\end{equation}

\noindent Here  $m_0$ is  a free  electron mass.  The calculations are
done for $\gamma  = 24 {\  \rm eV \AA^3}$  and $\alpha_0 =  7.2 {\ \rm
\AA^2}$. These  values allow  a good  description of  the experimental
data and are close to those obtained from $\mbox{\boldmath $k$}  \cdot
\mbox{\boldmath  $p$}$  and   tight-binding  calculations  for   ${\rm
Ga_{0.85}In_{0.15}As}$  (see  the  Appendix).  The  ratio $E_Z/E_F$ is
calculated    using     Fig.~\ref{EFermiPlot}.    When     calculating
$H^{(2)}_{SO}$, we have assumed that the average field in the well  is
one half of the maximum field ${\cal E} = 4 \pi e N_s/\kappa$. We have
also  taken  into  account  the  charge  in the depletion layer $N_0 =
0.58\cdot  10^{11}  {\  \rm  cm^{-2}}$.  The  value  of $\alpha_0$ was
calculated using Eq.~(\ref{kp}). If one takes into account  the
barriers, using theory of Refs.~\onlinecite{Gerchikov92,PikusUnp}  and
the self-consistently calculated wave functions, the value of $\alpha_0$
will increase by about 60\% for the electron densities in the interval
$N_s = (1-2)\cdot  10^{12} {\ \rm  cm^{-2}}$. This would  increase the
value of $H^{(2)}_{SO}$ approximately 2.5 times, but such large values
of $H^{(2)}_{SO}$  clearly do  not agree  with the  experiment. It  is
likely that the  barrier contribution depends  very strongly on  their
microscopic structure,  which may  be very  different from  the abrupt
interface model,  used in  the theory.  It is  also plausible that the
different barrier  structure is  responsible for  the relatively large
value of $H^{(2)}_{SO}$ for the sample {\sl C1}, for which $\alpha_0 =
8.8 {\ \rm \AA}$.

It   can   be   shown   using   the   Eq.~(\ref{Omega})  and  data  of
Fig.~\ref{EFermiPlot} that for $N_s <  N_{s0} = 7\cdot 10^{12} {\  \rm
cm^{-2}}$  the  Dresselhaus  term  decreases  with  increasing  $N_s$,
vanishes for $N_s = N_{s0}$, and then begins to increase. One can  see
from  Fig.~\ref{HSOPlot}  that  for  $N_s  >  1\cdot  10^{12}  {\  \rm
cm^{-2}}$ the Rashba term exceeds the Dresselhaus term.  Consequently,
we   denote   the   larger   contribution   in  Fig.~\ref{HSOPlot}  as
$H^{(2)}_{SO}$.


\mediumtext
\begin{table*}[t]
\caption{
Values  of  the  parameters  for  GaAs  and  InAs calculated using the
$sp_3s^*$  model  and  the  results  of  the  $k\cdot  p$  model.  The
parameters    of     $k\cdot    p$     model    were     taken    from
Ref.~\protect\onlinecite{Landolt},  except  those  marked by asterisk,
which were taken from Ref.~\protect\onlinecite{Hermann77}.  Parameters
for ${\rm Ga_{0.85}In_{0.15}As}$ were obtained by linear  interpolation
of the $k\cdot p$ model  parameters between GaAs and InAs.  The values
of $\gamma$ and $\alpha_0$ as obtained in these models are also given.
\label{ApTable}}
\begin{tabular}{ccccccc}
& \multicolumn{2}{c}{GaAs} & \multicolumn{2}{c}{InAs} &
\multicolumn{2}{c}{${\rm Ga_{0.85}In_{0.15}As}$} \\
\tableline
& $k\cdot p$ & $sp_3s^*$ & $k\cdot p$ & $sp_3s^*$ & $k\cdot p$ & $sp_3s^*$ \\
\tableline
$E_g$ (eV) & 1.519 & 1.5192 & 0.42 & 0.418 & 1.35 & 1.354 \\
$\Delta$ (eV) & 0.341 & 0.341 & 0.38 & 0.38 & 0.347 & 0.347 \\
$E_g^\prime$ (eV) & 2.97 & 2.98 & 3.97 & 3.95 & 3.12 & 3.104 \\
$\Delta^\prime$ (eV) & 0.171 & 0.159 & 0.24 & 0.26 & 0.181 & 0.20 \\
$P$ (eV \AA) & 10.49$^*$ & 10.23 & 9.2$^*$ & 9.22 & 10.29 & 10.16 \\
$P^\prime$ (eV \AA) & 4.78$^*$ & 1.46 & 0.87$^*$ & 1.06 & 4.20 & 1.03 \\
$Q$\tablenote{The sign of $Q$ in $k\cdot p$ model is not defined can be
chosen to be sthe same as in $sp_3s^*$ model.}
 (eV \AA) & -8.16$^*$ & -7.0 & -8.33$^*$ & -7.27 & -8.18 & -7.03 \\
$\frac{m}{m_0}$ & 0.0665 & 0.066 & 0.023 & 0.023 & 0.06 & 0.06 \\
$\gamma {\rm \ (eV \AA^3})$ & 27.5 & 10 & 26.9 & 71 & 27.7 & 13 \\
$\alpha_0 {\rm \ (\AA^2)}$ & 5.33 & 5.15 & 116.74 & 118.5 & 7.2
& 7.05 \\
\end{tabular}
\end{table*}
\narrowtext


One can see  from Figure~\ref{HSOPlot} that  the general character  of
the  density  dependence  of  $H^{(1)}_{SO}$ and $H^{(2)}_{SO}$ agrees
with the theory, and their values are close to those calculated  using
the above values of $\gamma$ and $\alpha_0$.

In Fig.~\ref{HSOPlot} b we show  a similar density dependence but  for
the  cubic  in  $k$  Dresselhaus  term  $H_{SO}  - H^\prime_{SO}$. The
theoretical formula for this field is

\begin{equation}
H_{SO} - H^\prime_{SO} = \eta_1 \gamma^2 \left(\frac{m}{m_0}\right)^2
N_s^2 \frac{\tau_3}{\tau_1}.
\end{equation}

\noindent The  top curve  corresponds to  $\tau_1/\tau_3 =  1$ and the
bottom one -- to $\tau_1/\tau_3 = 2$.

In the case of isotropic scattering, which is the case of short  range
potentials   scattering,   probabillity   $W(\varphi)$   in    formula
Eq.~(\ref{taun}) is angle independent and $\tau_1/\tau_3 = 1$. If only
small angle scattering is important (that is the case of scattering by
the   Coulomb   potential)   then   $\tau_1/\tau_3   =   9$  {\bf(}see
Eq.~(\ref{tauratio}){\bf)}. In our case we find $\tau_1/\tau_3$ to  be
in the range  from 1 to  2. It is  probably because scattering  in our
samples is the mixture of  short and long range scattering.  The short
range scattering is probably due to alloy scattering that is known  to
be mobility  limiting mechanism  in GaInAs  quantum wells.  Long range
scattering  is  most  probably  due  to  scattering  on  the   ionized
impurities  in  the  $\delta$-doped  layer.  Role of scattering by the
charged  impurities  in  the  $\delta$-doped  layer  was  confirmed by
observations      of      charge      correlation     effects     (see
Ref.~\onlinecite{Litwin94}).

\section{Conclusion}

In conclusion, we have presented new experimental studies of  positive
magnetoconductivity  caused  by  the  weak localization in selectively
doped  GaIn  As  quantum  wells  with different carrier densities. The
complete  interpretation  of  the  observations  is  obtained  in  the
framework  of  recently  developed  comprehensive  theory  of  quantum
corrections to conductivity.  In this theory,  we correctly take  into
account both linear  and cubic in  the wave vector  terms of the  spin
splitting  Hamiltonian.  These  terms  arise  due  to  the lack of the
inversion  symmetry  of  the  crystal.  We  also  include  the  linear
splitting  terms  which  appear  when  the  quantum well itself is not
symmetric.

It is shown that  in the density range  where all the above  terms are
comparable, the new theory allows  not only to achieve good  agreement
with  the  experiment  but,  unlike  earlier  theories, also gives the
values for the parameters of the spin splitting which are in agreement
with  previous  optical  experiments  \cite{Pikus84,Jusserand95}   and
theoretical calculations. Therefore, our research answers the question
what  spin  relaxation  mechanism  dominates  for
different electron densities and how it should be taken into account to
describe the weak localization and antylocalization phenomena in  quantum  wells  

\section{Acknowledgments}

S.  V.  Iordanskii  and  G.  E.  Pikus  thank  CNRS  and University of
Montpellier for invitation and  financial support during they  stay in
France. We would especially thank M.  I. Dyakonov and V. I. Perel  for
helpful advice and illuminating discussions. We would like also  thank
B. Jusserand, B. Etienne, T. Dietl for useful discussions. The authors
acknowledge support by the San Diego Supercomputer Center, where  part
of the calculations were performed. The research was supported in part
by the Soros  Foundation (G. E.  P.). F. G. P. acknowledges the  support by
NSF grant DMR993-08011  and by  the Center for  Quantized
Electronic Structures (QUEST) of  UCSB.  Authors affiliated
with   the   Universite   Montpellier   acknowledge  the  support  from
Schlumberger  Industries  and  Ministere  de  la  Recherche  et  de la
Technologie.

\appendix

\section{Spin Splitting in GaAs, InAs, and GaInAs}

Below  we  present  results  of  the  calculations  of  $\alpha_0$ and
$\gamma$   for   GaAs,   InAs   and  ${\rm  Ga_{0.85}In_{0.15}As}$  in
$\mbox{\boldmath $k$}  \cdot \mbox{\boldmath  $p$}$ and  tight-binding
calculations. The tight-binding calculations were done in the 20  band
tight-binding       model        including       the        spin-orbit
coupling\cite{Vogl83,Kobayashi82}.  Our  calculations  of   electronic
properties use $sp_3s^*$ tight-binding parameters especially chosen so
as to reproduce several features of the fundamental properties of bulk
constituents.  We  state  some  analytical  relations  connecting  the
effective masses and the deformation potentials at the $\Gamma$ point,
and  the  fifteen  parameters  of  the $sp_3s^*$ 20 band tight-binding
model\cite{Bertho}. Using these relations, as well as other  relations
between  the   fifteen  parameters   and  $\Gamma$   and  $X$   energy
values\cite{Kobayashi82}, we get a set of parameters which  accurately
reproduces  the  effective  masses  at  the  $\Gamma$ point, the [001]
deformation potential  and overall  band structure  in accordance with
reflectivity and photoemission measurements\cite{Landolt}.

Such a procedure has been  already checked to give a  good description
of    reflectivity    data    in    uniaxially   stressed   GaAs/${\rm
Ga_{0.89}In_{0.11}As}$ superlattices\cite{Boring92}.  In this  work we
use it  to obtain  InAs and  ${\rm Ga_{0.85}In_{0.15}As}$  parameters.
Using  these  parameters,  we  calculate  $\alpha_0$  on  the basis of
Eq.~(\ref{kp}).  In  order  to  determine  the  value  of  $\gamma$ we
calculate the value  of the spin  splitting as a  function of k  along
(110) direction. It  follows a cubic  dependence in $k$  from which we
extract  the  values  of  $\gamma$  given  in Table~\ref{ApTable}. The
parameters of  the $\mbox{\boldmath  $k$} \cdot  \mbox{\boldmath $p$}$
model  were   taken  from   Refs.~\onlinecite{Landolt,Hermann77}.  The
$\mbox{\boldmath $k$} \cdot \mbox{\boldmath $p$}$ parameters for ${\rm
Ga_{0.85}In_{0.15}As}$ were obtained  by linear interpolation  between
GaAs and InAs.

In the 3-band $\mbox{\boldmath $k$} \cdot \mbox{\boldmath $p$}$  model
one takes into account the states of the conduction band $\Gamma_1$  (
$\Gamma_6$)   with   the   Bloch   function   $S$,  the  valence  band
$\Gamma_{15\,v}$ ($\Gamma_8 + \Gamma_7$) with functions $X$, $Y$, $Z$,
and  the  higher  band  $\Gamma_{15\,c}$ ($\Gamma_{8c} + \Gamma_{7c}$)
with functions $X^\prime$, $Y^\prime$, and $Z^\prime$. The energies of
these  states  at  $k  =  0$  are: $E_{\Gamma_6} = 0$, $E_{\Gamma_8} =
-E_g$,  $E_{\Gamma_7}  =  -   (E_g  +  \Delta)$,  $E_{\Gamma_{7c}}   =
E^\prime_g$, and  $E_{\Gamma_{8c}} =  E^\prime_g +  \Delta^\prime$. In
this  model  $m_0/m$,  $\gamma$,  and  $\alpha_0$  are  given  by  the
following                                                  expressions
\cite{Pikus84,Cardona88,Ivchenko95,Malcher88,Rossler89,Gerchikov92,Note1}:

\begin{eqnarray}
\frac{m_0}{m} & = & 1 + \frac{2}{3} \frac{m_0}{\hbar^2} \Biggl\{
P^2 \frac{3E_g + 2\Delta}{E_g(E_g + \Delta)} +
P^{\prime^2} \frac{3 E_g^\prime + \Delta^\prime}{E^\prime_g
(E_g^\prime + \Delta^\prime)}\Biggr\},
\nonumber \\
\gamma & = & - \frac{4}{3} \frac{P P^\prime Q}{E_g(E_g^\prime +
\Delta^\prime)} \left(\frac{\Delta}{E_g + \Delta} +
\frac{\Delta^\prime}{E_g^\prime}\right),
\label{kp} \\
\alpha_0 & = & \frac{1}{3} \Biggl\{ P^2 \left[ E_g^{-2} -
(E_g + \Delta)^{-2} \right] -
\nonumber \\
& & P^{\prime^2} \left[ {E_g^\prime}^{-2} -
(E_g^\prime + \Delta^\prime)^{-2} \right] \Biggl\},
\nonumber
\end{eqnarray}

\noindent  where  $P  =  i  \hbar/  m_0  \,  \langle  S|p_z|Z\rangle$,
$P^\prime = i \hbar/ m_0 \, \langle S|p_z|Z^\prime\rangle$, and $Q = i
\hbar/ m_0  \, \langle  X |  p_y |  Z^\prime\rangle$ are the interband
matrix elements,  $m_0$ is  the free  electron mass,  $\mbox{\boldmath
$p$}  =  -i  \hbar  \nabla$.  Here  we  do  not  take into account the
contribution into $\gamma$ and $\alpha_0$ which arises from spin-orbit
mixing of the states $\Gamma_{15\,v}$ and $\Gamma_{15\,c}$.

The  values   of  $\gamma$   obtained  for   GaAs  from  tight-binding
calculations are usually  smaller then those  given by the  $k\cdot p$
model     (see      Table~\ref{ApTable}).     For      example,     in
Ref.~\onlinecite{Cardona94}  the  tight-binding  calculations give the
value  $2\gamma  =  17.8  {\  \rm  eV  \AA^3}$.  In  the  later   work
Ref.~\onlinecite{Cardona95}  for  the  same  parameter $2\gamma$ (this
time called $\gamma$) authors obtain the value $17 {\ \rm eV  \AA^3}$.
Calculated values of $\gamma$  should be compared to  the experimental
values $24 {\  \rm eV \AA^3}$  for bulk GaAs\cite{Pikus84}  and to the
recently      obtained      value      for     GaAs/GaAlAs     quantum
wells\cite{Jusserand95} $16.5 \pm 3 {\ \rm eV \AA^3}$.

\eject

\end{document}